\documentclass[12pt]{article}
\usepackage[dvips]{graphicx}
\usepackage{epsfig}

\def\nn{\nonumber}
\def\dfrac{\displaystyle\frac}
\def\etal{{\it et al.}}

\def\ie{{\it i.e.,~}}
\def\eg{{\it e.g.~}}

\def\JHEP#1#2#3{JHP {\bf #1}, #2 (#3)}
\def\PR#1#2#3{Phys. Rev. {\bf #1}, #2 (#3)}
\def\PRL#1#2#3{Phys. Rev. Lett. {\bf #1}, #2 (#3)}
\def\PL#1#2#3{Phys. Lett. {\bf #1}, #2 (#3)}

\def\NP#1#2#3{Nucl. Phys. {\bf #1}, #2 (#3)}

\def\PTP#1#2#3{Prog. Theor. Phys. {\bf #1}, #2 (#3)}
\def\EPJ#1#2#3{Eur. Phys. J. {\bf #1}, #2 (#3)}

\def\eqref#1{eq.~(\ref{eqn:#1})}
\def\Eqref#1{Equation~(\ref{eqn:#1})}
\def\eqsref#1{eqs.~(\ref{eqn:#1})}

\def\eqvref#1{~(\ref{eqn:#1})}

\def\eqlab#1{\label{eqn:#1}}

\def\bmaT{\left(\begin{array}{ccc}}
\def\emaT{\end{array}\right)}
\def\bma{\left( \begin{array} }
\def\ema{\end{array} \right)}

\def\wt{\widetilde}
\def\l{\left}
\def\r{\right}
\def\gsim{~{\rlap{\lower 3.5pt\hbox{$\mathchar\sim$}}\raise 1pt\hbox{$>$}}\,}
\def\lsim{~{\rlap{\lower 3.5pt\hbox{$\mathchar\sim$}}\raise 1pt\hbox{$<$}}\,}
\def\cU#1#2{U_{#1}^{#2}}
\def\cUm#1#2{\wt{U}_{#1}^{#2}}
\def\cA#1#2{{\cal A}_{#1}^{#2}}
\makeatletter
\newtoks\@stequation

\def\subequations{\refstepcounter{equation}%
  \edef\@savedequation{\the\c@equation}%
  \@stequation=\expandafter{\theequation}%   %only want \theequation
  \edef\@savedtheequation{\the\@stequation}% %expanded once
  \edef\oldtheequation{\theequation}%
  \setcounter{equation}{0}%
  \def\theequation{\oldtheequation\alph{equation}}}

\def\endsubequations{%
  \ifnum\c@equation < 2 \@warning{Only \the\c@equation\space subequation
    used in equation \@savedequation}\fi
  \setcounter{equation}{\@savedequation}%
  \@stequation=\expandafter{\@savedtheequation}%
  \edef\theequation{\the\@stequation}%
  \global\@ignoretrue}
\def\eqnarray{\stepcounter{equation}\let\@currentlabel\theequation
\global\@eqnswtrue\m@th
\global\@eqcnt\z@\tabskip\@centering\let\\\@eqncr
$$\halign to\displaywidth\bgroup\@eqnsel\hskip\@centering
     $\displaystyle\tabskip\z@{##}$&\global\@eqcnt\@ne
      \hfil$\;{##}\;$\hfil
     &\global\@eqcnt\tw@ $\displaystyle\tabskip\z@{##}$\hfil
   \tabskip\@centering&\llap{##}\tabskip\z@\cr}
\makeatother

\begin{document}
\renewcommand{\thefootnote}{\fnsymbol{footnote}}
\renewcommand{\thefigure}{\arabic{figure}}
\title{%
Effect of the smaller mass-squared difference\\
for the long base-line neutrino experiments
}
\author{
{Naotoshi Okamura}\thanks{E-mail address : okamura@yukawa.kyoto-u.ac.jp} \\
{\small \it Korea Institute for Advanced Study}\\
{\small \it 207-43 Cheongnyangi 2-dong, Dongdaemun-gu, Seoul 130-722,
Korea}\\
{\small \it Yukawa Institute for Theoretical Physics}\\
{\small \it Kyoto University, Kyoto 606-8502, Japan}\footnote{{%
after May 1st, 2005.
}}
}
\date{}
\maketitle
\vspace{-9.5cm}
\begin{flushright}
KIAS-P04045
\hspace*{3ex}
hep-ph/0411388
\end{flushright}
\vspace{ 9.5cm}
\vspace{-2.0cm}
\begin{abstract}
Usually, neutrino oscillation experiments are analyzed
within the two-flavor framework
which is governed by
1 mass-squared difference and 1 mixing angle.
But there are 6 parameters,
2 mass-squared differences,
3 mixing angles,
and
1 CP phase
within the three-flavor framework.
In this article,
we estimate the effect from the smaller mass-squared difference,
the other mixing angles, and the CP phase,
which we call three-flavor effect, 
for the determination of
the mass-squared difference and the mixing angle
from the $\nu_\mu$'s survival and transition probability
with the two-flavor analysis.
It is found that
the mass-squared difference from the 
two-flavor analysis
is slightly shifted from 
the larger mass-squared difference
by the three-flavor effect.
The order of magnitude of the three-flavor effect
for the mass-squared difference
is comparable with that of the expected error
for the mass-squared difference of the two-flavor analysis
in the future long base-line
neutrino oscillation experiments.
The CP phase dependence of the $\nu_\mu\to\nu_e$
transition probability is also shown.
\end{abstract}

\section{introduction}
 The three neutrino framework has 9 physical parameters:
3 neutrino masses,
3 mixing angles,
and
3 CP violating phase,
if neutrinos are Majorana particles.
Neutrino oscillation experiments are sensitive to 6 parameters:
2 mass-squared difference,
3 mixing angles,
and 
1 CP phase.
Usually, data from the experiments are analyzed
within the two-flavor framework, which is governed by only
1 mass-squared difference and 1 mixing angle.
So far, 
for the long base-line
neutrino oscillation experiments,
we have been able to neglect the effect from
the smaller mass-squared difference, 
the other mixing angles, and CP phase,
which we call three-flavor effect.
This is because 
the error of the larger mass-squared difference,
related to the atmospheric neutrino observations~\cite{atm}
and the on-going long base-line experiment K2K~\cite{K2K},
is larger than the smaller mass-squared difference,
which is obtained from solar neutrino observations~\cite{solar} and
the reactor neutrino experiment KamLAND~\cite{KamLAND}.

The long base-line neutrino oscillation experiments
in future~\cite{T2K,nova,BNL} 
plan to measure the mass-squared difference
and mixing parameter precisely.
Because the order of magnitude of the ratio between
the smaller mass-squared difference and the larger one
is supposed to be similar to that of expected
error of the future long base-line experiments,
it is necessary to take into account
the contribution of the three-flavor effect
in 
the determination of the mass-squared difference
and mixing angle from the future long base-line 
precision measurements.
In this article, we estimate this effect
using the three-flavor framework.
We discard the result of the LSND experiment~\cite{LSND}.
Obviously, this analysis is, in general, not valid,
if the LSND result is confirmed by the
MiniBooNE experiment~\cite{miniboone}.
From the survival probability of $\nu_\mu$,
we find that the larger mass-squared difference is shifted by the
three-flavor effect
and that
the order of magnitude of this shift
depends on the neutrino energy
and is similar to 
that of the smaller mass-squared difference.
We obtain the same result from the transition probability
$\nu_\mu^{} \to \nu_e^{}$,
and also find the CP phase dependence for the
transition probability.
 A lot of groups have analyzed the experimental data with
the three-flavor framework numerically \cite{nume}.
 However, these analyses cannot point out the specific reason
for the value of parameters.
 In this paper, we point out the specific contribution to the
parameters from the three-flavor effect.
 We think that 
these formulations are useful to study in the numerical analysis.

 This article is organized as follows.
 In section 2, we will show the useful notations and 
the convenient form of the probability for easy estimating
the contribution of the three-flavor effect.
 In section 3, we will estimate the three-flavor effect for the
$\nu_\mu$ disappearance mode.
 We will also estimate the three-flavor effect for the
$\nu_\mu \to \nu_e$ transition mode, in section 4.
 Finally, we will be devoted to the summary in the last section.

\section{notations}
 In the three neutrino framework, 
and in the basis in which the charged leptons are diagonal,
the three weak interaction
eigenstates, $\nu_\alpha^{}$ ($\alpha = e,\mu,\tau$) are
expressed as 
\begin{equation}
 \nu_\alpha^{} =
  \sum^{3}_{i=1} (V_{\rm MNS})_{\alpha i}~\nu_i^{}\,,
\end{equation}
where $\nu_i^{}$ are the three mass eigenstates and $V_{\rm MNS}$ is
the Maki-Nakagawa-Sakata (MNS) matrix~\cite{MNS}.
We adopt the following parameterization~\cite{PDB}
\begin{equation}
 V_{\rm MNS}^{} = U_{}^{} {\cal P} 
  = U_{}^{}~{\rm diag}(e^{i\alpha_1/2},e^{i\alpha_2/2},1)\,,
\end{equation}
where ${\cal P}$ cannot be determined from the neutrino oscillation
experiment.
 The matrix $U$, which has three mixing angles and one phase,
can be parameterized in the same way as the Cabibbo-Kobayashi-Maskawa 
matrix~\cite{CKM}.
Because the present neutrino oscillation experiments
constrain directly the elements
$U_{e2}$, $U_{e3}$, and $U_{\mu 3}$,
we find it most convenient to adopt the parameterization~\cite{HO1,H2B},
where these three elements in the upper-right
corner of the $U$ matrix are the independent parameters.
Without losing generality, we can take 
$U_{e2}$ and $U_{\mu 3}$ to be real and non-negative.
By allowing $U_{e3}$ to have the complex phase,
\begin{equation}
U_{e2}\,,U_{\mu 3} \geq 0,~~~ 
U_{e3}\equiv \left|U_{e3}\right| e^{-i\phi} 
~~(0 \leq \phi < 2\pi )\,,
\end{equation}
these $U_{e2}, U_{\mu 3}, |U_{e3}|$, and $\phi$
are the four independent parameters.
All the other elements of $U$
are then determined by the unitary
conditions,
\begin{subequations}
\begin{eqnarray}
U_{e1} &=& \sqrt{1-|U_{e3}|^2-|U_{e2}|^2}\,, 
~~~~~U_{\tau 3}
=
\sqrt{1-|U_{e3}|^2-|U_{\mu 3}|^2}\,, 
\\ 
U_{\mu 1} &=& - \frac{U_{e2}U_{\tau 3} + U_{\mu 3}U_{e1}U_{e3}^{\ast} }
                      {1-|U_{e3}|^2}\,, 
~~U_{\mu 2} = \frac{U_{e1}U_{\tau 3} - U_{\mu 3}U_{e2}U_{e3}^{\ast} }
                    {1-|U_{e3}|^2}\,,\\
U_{\tau 1} &=& \frac{U_{e2}U_{\mu 3} - U_{\tau 3}U_{e1}U_{e3}^{\ast} }
                     {1-|U_{e3}|^2}\,,
~~~~U_{\tau 2} = - \frac{U_{\mu 3}U_{e1} + U_{e2}U_{\tau 3}U_{e3}^{\ast} }
                       {1-|U_{e3}|^2}\,.
\end{eqnarray}
\eqlab{def_MNS}
\end{subequations}
For the convenience, the independent parameters in the MNS matrix
are rewritten as
\begin{equation}
U_{e3}^{} \equiv \sin \theta_{13}\,,
~~~~~~
U_{e2}^{} \equiv \sin \theta_{12} \cos \theta_{13}\,,
~~~~~~
U_{\mu3}^{} \equiv \sin \theta_{23} \cos \theta_{13}\,.
\eqlab{another_MNS}
\end{equation}

The atmospheric neutrino oscillation experiments, which measure the $\nu_\mu$
survival probability determine the absolute values of the larger
mass-squared differences and one-mixing angle \cite{atm} as
\begin{equation}
1.5\times 10^{-3}<
\left| \delta m^2_{\rm atm} \right|
< 3.4 \pm 0.5 \times 10^{-3} \mbox{{eV}}^2\,,
~~~~\mbox{{and}}~~~
\sin^22\theta_{\rm atm} >0.92 \,,
\eqlab{atm}
\end{equation}
at the 90\% C.L.
The K2K experiment \cite{K2K} confirms the above results. 
These values are planed to measure more precisely,
a few percent order by the future long base-line experiments
\cite{T2K,nova,BNL}.
The solar neutrino experiments, which measure the $\nu_e$ survival
probability in the sun \cite{solar},
and the KamLAND experiment which measure the $\bar{\nu}_e$ survival
probability from the reactors \cite{KamLAND},
determine the smaller mass-squared difference and another mixing angle
as
\begin{equation}
 \delta m^2_{\rm sol} = 8.2 ^{+0.6}_{-0.5} \times 10^{-5} \mbox{{eV}}^2\,,
~~~~\mbox{{and}}~~~
\tan^2\theta_{\rm sol} = 0.40 ^{+0.09}_{-0.07}\,.
\eqlab{sol}
\end{equation}
It is remarkable point that the order of the smaller mass
squared-difference is as same as that of expected error of the future
long base-line experiments \cite{T2K,nova,BNL}.
Thus, it is necessary to take into account the contribution of the
three-flavor effect in the determination of the mass-squared difference
and mixing angle analytically,
because experimental data are analyzed ordinary 
in the two-flavor framework.
The CHOOZ reactor experiment \cite{CHOOZ} gives the upper bound of the
third mixing angle as
\begin{eqnarray}
 \sin^22\theta_{\rm rct} &<& 0.20 
~~~\mbox{{for}}~~~\delta m^2 = 2.0 \times 10^{-3} \mbox{{eV}}^2\,,\nn\\
 \sin^22\theta_{\rm rct} &<& 0.16 
~~~\mbox{{for}}~~~\delta m^2 = 2.5 \times 10^{-3} \mbox{{eV}}^2\,,\nn\\
 \sin^22\theta_{\rm rct} &<& 0.14 
~~~\mbox{{for}}~~~\delta m^2 = 3.0 \times 10^{-3} \mbox{{eV}}^2\,,
\eqlab{rct}
\end{eqnarray}
at the 90\% C.L.
Since we can always take
$\left| \Delta_{12} \right| < \left| \Delta_{13} \right|$
without loosing generality, 
we assume that
$\Delta_{12}$ is from the results of solar neutrino and reactor
anti-neutrino observations
and 
$\Delta_{13}$ is related to the atmospheric and long base-line
neutrino experiments.
The sign of $\Delta_{12}$ is determined form the solar neutrino
observation, $\Delta_{12} > 0$.
However, that of $\Delta_{13}$ cannot be determined by any
observations.
In this article, we call $\Delta_{13} > 0$ ``normal hierarchy''
and $\Delta_{13} < 0$ ``inverted hierarchy''.
Under the 
$\left| \Delta_{12} \right| < \left| \Delta_{13} \right|$ relation,
$\theta_{\rm atm}$, $\theta_{\rm sol}$, and 
$\theta_{\rm rct}$ are related to the MNS matrix elements as :
\begin{eqnarray}
2 \cU{\mu 3}{2} &=& 
1-\sqrt{1-\sin^2 2\theta_{\rm atm}}\,,\nn\\
2 \left|\cU{e3}{}\right|^2 &=& 
1-\sqrt{1-\sin^2 2\theta_{\rm rct}}\,,\nn\\
2\cU{e2}{2} &=& 
1-|\cU{e3}{}|^2 -
\sqrt{\left(1-|\cU{e3}{}|^2\right)^2-\sin^22\theta_{\rm sol}}
\,.
\end{eqnarray}

Starting from the flavor eigenstate $\alpha$, 
the probability of finding the
flavor eigenstate $\beta$
at the base-line length $L$
is, in vacuum,
\begin{subequations}
\begin{eqnarray}
P_{\nu_\alpha^{} \to \nu_\beta^{}} &=&
\left|\sum_{j=1}^3 (V_{\rm MNS})_{\beta j}^{} 
\exp\left(-i\frac{m_j^2}{2E_\nu}L\right)
(V_{\rm MNS})_{\alpha j}^{\ast} 
\right|^2\eqlab{P_ex_00} \\
&=&
\delta_{\alpha\beta}
-4{\rm Re}\left\{
\cU{\alpha 3}{}\cU{\beta 3}{\ast}\cU{\beta 1}{}\cU{\alpha 1}{\ast}
+
\cU{\alpha 2}{}\cU{\beta 2}{\ast}\cU{\beta 3}{}\cU{\alpha 3}{\ast}
\right\}
\sin^2\dfrac{\Delta_{13}^{}}{2}\nn\\
&&~~~~
-4{\rm Re}\left\{
\cU{\alpha 2}{}\cU{\beta 2}{\ast}\cU{\beta 1}{}\cU{\alpha 1}{\ast}
+
\cU{\alpha 2}{}\cU{\beta 2}{\ast}\cU{\beta 3}{}\cU{\alpha 3}{\ast}
\right\}
\sin^2\dfrac{\Delta_{12}^{}}{2}\nn\\
&&~~~~
+2{\rm Re}
 \left(
 \cU{\alpha 2}{}\cU{\beta 2}{\ast}\cU{\beta 3}{}\cU{\alpha 3}{\ast} 
 \right)
\left(
\sin\Delta_{12}\sin\Delta_{13}
+ 4\sin^2\dfrac{\Delta_{12}}{2}\sin^2\dfrac{\Delta_{13}}{2}
\right)
\nn\\
&&~~~~
- 4J_{\rm MNS}^{(\alpha,\beta)}\left(
 \sin^2\dfrac{\Delta_{13^{}}}{2}\sin\Delta_{12} -
 \sin^2\dfrac{\Delta_{12^{}}}{2}\sin\Delta_{13}
\right)\,,
\eqlab{P_ex_03}
\end{eqnarray}
\eqlab{P_ex_0}
\end{subequations}
where $J_{\rm MNS}^{(\alpha,\beta)}$ is the 
Jarlskog parameter~\cite{JarP}:
\begin{eqnarray}
J_{\rm MNS}^{(\alpha,\beta)}~
&\equiv&~{\rm Im}
\left((V_{\rm MNS})_{\alpha i}
      (V_{\rm MNS})_{\beta i}^{\ast}
      (V_{\rm MNS})_{\beta j} 
      (V_{\rm MNS})_{\alpha j}^{\ast}\right) \nn\\
&=& {\rm Im} \left({U_{\alpha i} U_{\beta i}^{\ast} U_{\beta j} 
U_{\alpha j}^{\ast}}\right)
= - \frac{U_{e1}U_{e2}U_{\mu 3}U_{\tau 3}}{1-\left|U_{e3}\right|^2}
              {\rm Im}\left({U_{e3}}\right)
~\equiv~A \sin \phi\,,
\eqlab{Jmns}
\end{eqnarray}
which is defined to be positive 
for $(\alpha,\beta)=(e,\mu)$, $(\mu,\tau)$, $(\tau,e)$
and $(i,j)=(1,2),(2,3),(3,1)$.
$A$ is the absolute value of the Jarlskog parameter.
In addition, $\Delta_{ij}$ is
\begin{equation}
\Delta_{ij}
\equiv
\dfrac{m^2_j - m^2_i}{2E_\nu}L
=
\dfrac{\delta m_{ij}^2}{2E_\nu}L
\simeq
 2.534 \dfrac{\delta m_{ij}^2 ({\rm eV}^2)}{E_\nu({\rm GeV})}L({\rm km}
)\,,
\end{equation}
where $E_\nu$ is the neutrino energy.

We rewrite \eqref{P_ex_0} in a form convenient to estimate the
contribution of the smaller mass-squared difference,
\begin{equation}
P_{\nu_\alpha^{} \to \nu_\beta^{}} 
\equiv
P_0(\alpha,\beta) + P_1(\alpha,\beta) \times \sin \Delta_{12}^{}
+ P_2(\alpha,\beta) \times 4\sin^2\dfrac{\Delta_{12}}{2}\,,
\eqlab{P_ex_11}
\end{equation}
where each term is
\begin{subequations}
\begin{eqnarray}
 P_0(\alpha,\beta) &=&
\delta_{\alpha\beta}-4{\rm Re}\left\{
\cU{\alpha 3}{}\cU{\beta 3}{\ast}\cU{\beta 1}{}\cU{\alpha 1}{\ast}
+
\cU{\alpha 2}{}\cU{\beta 2}{\ast}\cU{\beta 3}{}\cU{\alpha 3}{\ast}
\right\}
\sin^2\dfrac{\Delta_{13}^{}}{2}\,, 
\eqlab{P_ex_p0}\\
P_1(\alpha,\beta) &=&
2{\rm Re}
 \left(
 \cU{\alpha 2}{}\cU{\beta 2}{\ast}\cU{\beta 3}{}\cU{\alpha 3}{\ast} 
 \right)
 \sin\Delta_{13}^{}
 -
 4J_{\rm MNS}^{(\alpha,\beta)} \sin^2\dfrac{\Delta_{13^{}}}{2}\,,
\eqlab{P_ex_p1}
\\
P_2(\alpha,\beta) &=&
-
{\rm Re}\left(
 \cU{\alpha 1}{}\cU{\beta 1}{\ast}\cU{\beta 2}{}\cU{\alpha 2}{\ast} 
+
 \cU{\alpha 2}{}\cU{\beta 2}{\ast}\cU{\beta 3}{}\cU{\alpha 3}{\ast} 
\cos \Delta_{13}^{}
\right)
+ J_{\rm MNS}^{(\alpha,\beta)} \sin\Delta_{13}^{}\,.
\eqlab{P_ex_p2}
\end{eqnarray}
\eqlab{P_ex_p}
\end{subequations}

All the above formulas remain valid 
when replacing the mass-squared differences and
the MNS matrix elements with the ``effective'' ones,
\begin{equation}
\Delta_{ij} \to \wt \Delta_{ij}\,,
~~~
U_{\alpha i} \to \wt U_{\alpha i}\,,
~~~
J_{\rm MNS}^{(\alpha,\beta)} \to \wt J_{\rm MNS}^{(\alpha,\beta)}\,,
\end{equation}
as long as the matter density remains the same along the base-line.
The effective parameters 
$\wt U_{\alpha i}$ are
defined from the following Hamiltonian
\begin{equation}
{{\cUm{}{}}}
\bmaT
 {\lambda_1} & 0 & 0 \\
 0 & {\lambda_2} & 0 \\
 0 & 0 & {\lambda_3}
\emaT
{{\cUm{}{\dagger}}}
=
{\cU{}{}}
\bmaT
 0 & 0 & 0 \\
 0 & \delta m^2_{12} & 0 \\
 0 & 0 & \delta m^2_{13}
\emaT
{\cU{}{\dagger}}
+ 
\bmaT
a & 0 & 0 \\
0 & 0 & 0 \\
0 & 0 & 0
\emaT\,,
\eqlab{H_mat}
\end{equation}
and $\wt \Delta_{ij}$ is defined as
\begin{equation}
\wt{\Delta}_{ij}
\equiv
\dfrac{\delta \wt{m}_{ij}^{2}}{2E_\nu}L
=
\dfrac{{\lambda_j}-{\lambda_i}}{2E_\nu}L\,.
\end{equation}
The term $a$ in \eqref{H_mat}
stands for the matter effect~\cite{MSW},
\begin{eqnarray}
a(E_\nu^{})
=2\sqrt{2}G_F n_e E_\nu^{} 
={7.56}\times 10^{-5}({\rm eV}^2)\left(\frac{\rho}{{\rm g/cm}^{3}}\right)
  \left(\frac{E_\nu}{\rm GeV}\right)\,,
\eqlab{matter_a}
\end{eqnarray}
where
$n_e$ is the electron density of the matter,
$G_F$ is the Fermi constant,
and $\rho$ is the matter density.
When $\delta m^2_{12} < a < \delta m^2_{13}$
and $ U_{e3}^{} \ll O(1)$,
the effective mass-squared differences
are written as
\begin{equation}
\wt{\Delta}_{13} \simeq \Delta_{13}^{}-\Delta_{12}^{}\cos^2\theta_{12}^{}\,,
~~~~~
\wt{\Delta}_{12} \simeq  \dfrac{a}{2E}L-\Delta_{12}^{}\cos2\theta_{12}^{}\,,
\eqlab{ele_app1}
\end{equation}
and the mixing angles become
\begin{equation}
\wt{\theta}_{23}^{} \simeq \theta_{23}\,,
~~~~~
\wt{\theta}_{13}^{} \simeq \l(1+\dfrac{a}{\delta m^2_{13}}\r)
\theta_{13}^{}\,,
~~~~~
\tan 2 \wt{\theta}_{12}^{} \simeq 
\dfrac{\delta m^2_{12} \sin 2\theta_{12}^{}}
{\delta m^2_{12} \cos 2 \theta_{12}^{}- a}\,.
\eqlab{ele_app2}
\end{equation}
Hereafter, we use $U_{\alpha i}^{}$ and $\Delta_{ij}$ instead
of $\wt U_{\alpha i}^{}$ and $\wt \Delta_{ij}$, because of
simplicity.
 But we have to keep in mind that these values depend on the
neutrino energy.

\section{survival probability of $\nu_\mu$}
From \eqsref{P_ex_11} and\eqvref{P_ex_p},
the survival probability of $\nu_\mu^{}$
is written as
\begin{eqnarray}
P_{\nu_\mu^{} \to \nu_\mu^{}}=
P_0(\mu,\mu)
+ P_1(\mu,\mu) \times \sin{\Delta_{12}}
+ P_2(\mu,\mu) \times 4\sin^2\dfrac{\Delta_{12}}{2}\,,
\eqlab{Pmm}
\end{eqnarray}
where
\begin{subequations}
\begin{eqnarray}
P_0(\mu,\mu) &=& 1-4\left|\cU{\mu 3}{}\right|^2
\left(1-\left|\cU{\mu 3}{}\right|^2\right)
\sin^2\dfrac{\Delta_{13}}{2}\,,
\eqlab{Pmm0}\\
P_1(\mu,\mu)&=& 
2\left|\cU{\mu 2}{}\right|^2\left|\cU{\mu 3}{}\right|^2
\sin\Delta_{13}\,,
\eqlab{Pmm1}\\
P_2(\mu,\mu)&=&
-\left|\cU{\mu 2}{} \right|^2
\left(\left|\cU{\mu 1}{}\right|^2
+\left|\cU{\mu 3}{}\right|^2 \cos{\Delta_{13}}
\right)\,.
\eqlab{Pmm2}
\end{eqnarray}
\eqlab{Pmm_ex}
\end{subequations}
The survival probability of $\nu_\mu^{}$ in the two-flavor framework is
written as
\begin{equation}
 P_{\nu_\mu^{}\to\nu_\mu^{}}^{(2)}
=
1 - \sin^22\theta_{\mu\mu} \sin^2 \dfrac{\Delta_{\mu\mu}}{2}\,,
\eqlab{Pmm2f}
\end{equation}
where $\theta_{\mu\mu}$ is the mixing angle
and
$\Delta_{\mu\mu}$ is the mass-squared difference,
which are obtained from the two-flavor analysis.
We expect that all these parameters, especially $\Delta_{\mu \mu}$,
to be shifted by the three-flavor effect.
In order to estimate this effect,
we rewrite them as 
\begin{equation}
\sin^22\theta_{\mu\mu} = 1.0 - \varepsilon\,,
\hspace*{5ex} 
\Delta_{\mu\mu}       = \Delta_{13} + 2\delta \,,
\eqlab{app1}
\end{equation}
where $\delta$ denotes the three-flavor effect,
and $\varepsilon$ stands for the difference from 
the maximal mixing.
Both of them depend on the neutrino energy,
base-line length, and the oscillation parameters.
We assume that $\delta$ is smaller than
$\Delta_{13}$ and $\Delta_{13}\sim O(1)$ for long base-line experiments.

From the atmospheric neutrino observations and K2K experiment,
we already know that $\nu_\mu^{}$ oscillate to 
another flavor maximally at the first-dip,
$\Delta_{\mu\mu}(E_\nu=E_{\rm dip})=\pi$
for normal hierarchy.
By using \eqref{app1},
\ie $\Delta_{13} = \pi-2\delta^{\rm dip}$,
\eqref{Pmm_ex} becomes
\begin{subequations}
\begin{eqnarray}
P_0(\mu,\mu) &=& 1-4\left|\cU{\mu 3}{\rm dip}\right|^2
\left(1-\left|\cU{\mu 3}{\rm dip}\right|^2\right)
\cos^2\delta^{\rm dip} \,,
\eqlab{Pmmdip0}\\
P_1(\mu,\mu)&=& 
2\left|\cU{\mu 2}{\rm dip}\right|^2\left|\cU{\mu 3}{\rm dip}\right|^2
\sin2\delta^{\rm dip}\,,
\eqlab{Pmmdip1}\\
P_2(\mu,\mu)&=&
-\left|\cU{\mu 2}{\rm dip} \right|^2
\left(\left|\cU{\mu 1}{\rm dip}\right|^2
-\left|\cU{\mu 3}{\rm dip}\right|^2 \cos{2\delta^{\rm dip}}
\right)\,,
\eqlab{Pmmdip2}
\end{eqnarray}
\eqlab{Pmmdip_ex}
\end{subequations}
and \eqref{Pmm2f}
\begin{equation}
 P_{\nu_\mu^{}\to\nu_\mu^{}}^{(2)}
= \varepsilon^{\rm dip}\,,
\eqlab{Pmmdip2f}
\end{equation}
where 
the label ``dip'' in $|\cU{\mu i}{}|$,
$\delta$, and $\varepsilon$ means 
that these quantities take some fixed value
at the first-dip energy $E_{\rm dip}$.
From \eqsref{Pmmdip_ex} and\eqvref{Pmmdip2f},
we obtain
\begin{eqnarray}
 \delta^{\rm dip} 
 \simeq
\dfrac
{4\left|\cU{\mu3}{\rm dip}\right|^2
  \left(1-\left|\cU{\mu3}{\rm dip}\right|^2\right)
-\left(1^{}_{}-\varepsilon^{\rm dip}_{} \right)
}
{4\Delta_{12}^{\rm dip}\left|\cU{\mu 2}{\rm dip}\right|^2
\left|\cU{\mu 3}{\rm dip}\right|^2}
+
\dfrac
{ 1 
 - \left|\cU{\mu 2}{\rm dip}\right|^2
 -2\left|\cU{\mu 3}{\rm dip}\right|^2}
{4\left|\cU{\mu 3}{\rm dip}\right|^2}
\Delta_{12}^{\rm dip}\,.
\eqlab{dip_ex1}
\end{eqnarray}
The first term of \eqref{dip_ex1} has to be zero
because of the assumption $O(\delta) < 1$,
and therefore we obtain
\begin{equation}
\left|\cU{\mu 3}{\rm dip}\right|^2
=\dfrac{\left(1 \pm \sqrt{\varepsilon^{\rm dip}}\right)}{2}\,.
\eqlab{cond_dip1}
\end{equation}
When we take a negative sign in \eqref{cond_dip1},
$\delta^{\rm dip}$ becomes
\begin{equation}
\delta^{\rm dip}
\simeq
-\dfrac{\Delta_{12}^{\rm dip}}{2}
\left\{
\left|\cU{\mu 2}{\rm dip}\right|^2
+
\left(1-\left|\cU{\mu 2}{\rm dip}\right|^2\right)
\sqrt{\varepsilon^{\rm dip}}
-
\varepsilon^{\rm dip}
\right\}\,.
\end{equation}
Since the best fit value of the
mixing angle is maximum from the experiments~\cite{atm,K2K},
we take $\varepsilon^{\rm dip} = 0$.
Thus, $\delta^{\rm dip}$ simplifies to
\begin{eqnarray}
 \delta^{\rm dip}
 \simeq -\dfrac{\Delta_{12}^{\rm dip}}{2}
\left|\cU{\mu 2}{\rm dip}\right|^2\,.
\eqlab{dip1}
\end{eqnarray}
From \eqref{app1}, the larger mass-squared difference
at $E_{\rm dip}$ is
\begin{equation}
\delta m^2_{13}(E_{\rm dip})
=
\delta m^2_{\rm dip}
+
\delta m^2_{12}(E_{\rm dip}) \left|\cU{\mu 2}{\rm dip}\right|^2\,,
\eqlab{dip2}
\end{equation}
where 
\begin{equation}
 \delta m^2_{\rm dip} \equiv \dfrac{2\pi E_{\rm dip}}{L}\,,
\end{equation}
is from the two-flavor analysis.
\Eqref{dip2} denotes that 
the order of magnitude of the three-flavor effect is roughly
the same as that of the expected error of future experiments.
Since we know that both $\delta m^2_{12}$
and $\left|\cU{\mu 2}{\rm dip}\right|^2$ are positive,
the mass-squared difference from two-flavor analysis
is slightly smaller than the larger mass-squared difference.
We also obtain the relation between $\delta$ and 
the smaller mass-squared difference
at the first peak $E_{\rm peak}$,
where $\Delta_{\mu\mu}(E_{\rm peak})=2\pi$,
\begin{eqnarray}
\delta^{\rm peak}
\simeq - \dfrac{\Delta_{12}^{\rm peak}}{2}
\left(1-\left|\cU{\mu 2}{\rm peak}\right|^2\right)
\,.
\eqlab{peak1}
\end{eqnarray}
Thus, the larger mass-squared difference at $E_{\rm peak}$ is
\begin{equation}
\delta m^2_{13}(E_{\rm peak})
=
\delta m^2_{\rm peak}
+
\delta m^2_{12}(E_{\rm peak})
\left( 1- \left|\cU{\mu 2}{\rm peak}\right|^2\right)\,,
\eqlab{peak2}
\end{equation}
where 
\begin{equation}
 \delta m^2_{\rm peak} \equiv \dfrac{4\pi E_{\rm peak}}{L}\,,
\end{equation}
is also from the two-flavor analysis.
We obtain the same results for the inverted hierarchy.
Since $\delta m^2_{13}$ is not changed by matter effect at
$E_{\nu}<$10 GeV,
we obtain the relation between \eqref{dip2} and \eqref{peak2}:
\begin{equation}
\delta m^2_{\rm dip}
+
\delta m^2_{12}(E_{\rm dip}) \left|\cU{\mu 2}{\rm dip}\right|^2
=
\delta m^2_{\rm peak}
+
\delta m^2_{12}(E_{\rm peak})
\left( 1- \left|\cU{\mu 2}{\rm peak}\right|^2\right)\,.
\eqlab{rel_dip_peak}
\end{equation}
By using the definition of $\delta m^2_{\rm dip, peak}$,
we find
\begin{equation}
\dfrac{L}{2\pi}
 \dfrac{
\delta m^2_{12}\left(E_{\rm dip}\right)\left|\cU{\mu2}{\rm dip}\right|^2
-
\delta m^2_{12}\left(E_{\rm peak}\right)
\left(1-\left|\cU{\mu2}{\rm peak}\right|^2\right)
}
{2E_{\rm peak} - E_{\rm dip}}
=\pm 1\,,
\eqlab{IorIII}
\end{equation}
where the sign of the right-hand side corresponds to the type of
the mass hierarchy,
the positive sign being that of the normal hierarchy.
 From the \eqsref{dip2} and \eqvref{peak2}, we easily understand 
that the three-flavor effect for the larger mass-squared difference
depends on the energy. 
 When $\rho = 2.5$(g/cm$^3$) and $E_\nu^{}\simeq O(1)$GeV,
$\wt{\theta}_{12}^{}$ is to shift away from $\theta_{12}^{}$
towards $90^\circ$, which is obtained from \eqref{ele_app1}.
From \eqsref{def_MNS} and \eqvref{another_MNS}, 
the value of $|\cU{\mu 2}{}|$ becomes $0$.
Because $\delta m^2_{12}(E_{\nu})$ is also changed by the matter effect,
which is shown in \eqref{ele_app1},
\eqref{IorIII} becomes 
\begin{equation}
\dfrac{L}{2\pi}
\dfrac{
a(E_{\rm peak}) - \delta m^2_{12} \cos 2\theta_{12}^{}
}
{E_{\rm dip}-2E_{\rm peak}}
=\pm 1\,,
\eqlab{IorIII-2}
\end{equation}
where the value of $\delta m^2_{12}$ and $\theta_{12}^{}$ is
that of vacuum one, which are listed in \eqref{sol}.
This relation suggest us that 
we can determine the sign of the $\delta m^2_{13}$ from the 
long base-line experiments, 
when we measure the energy of ``peak'' and ``dip'' precisely
and we know the value of the smaller mass-squared difference,
mixing angle with small errors.
 This result cannot be obtained from the numerical analysis.
 This fact points out that we can pick up the three-flavor effect
from the fitting function of the survival probability which is
obtained from the experimental data.

\section{transition probability}
From \eqsref{P_ex_0} and\eqvref{P_ex_p}, the transition probability,
$\nu_\mu^{} \to \nu_e^{}$ is written as
\begin{eqnarray}
P_{\nu_\mu^{} \to \nu_e^{}}&=&
P_0(\mu,e)
+ P_1(\mu,e) \times \sin{\Delta_{12}}
+ P_2(\mu,e) \times 4\sin^2\dfrac{\Delta_{12}}{2}\,,
\eqlab{Pme}
\end{eqnarray}
where
\begin{subequations}
\begin{eqnarray}
P_0(\mu,e) &=&
4\left|\cU{\mu 3}{}\cU{e 3}{}\right|^2
\sin^2\dfrac{\Delta_{13}}{2}\,,
\eqlab{Pme0}\\
P_1(\mu,e) &=& 
2\left\{
{\rm Re}\left(
\cU{e 2}{\ast}\cU{\mu 2}{}
\cU{e 3}{}\cU{\mu 3}{\ast}
\right)
\sin\Delta_{13}
+
2J_{\rm MNS}^{(\mu,e)} \sin^2\dfrac{\Delta_{13}}{2}
\right\}\,,\\
P_2(\mu,e)&=&
-{\rm Re}\left(
\cU{\mu 1}{}\cU{e 1}{\ast}\cU{e 2}{}\cU{\mu 2}{\ast}
+
\cU{\mu 2}{}\cU{e 2}{\ast}\cU{e 3}{}\cU{\mu 3}{\ast}
\cos \Delta_{13}
\right)
-
J_{\rm MNS}^{(\mu,e)} \sin \Delta_{13}\,.
\eqlab{Pme2}
\end{eqnarray}
\eqlab{Pme_ex}
\end{subequations}
Here,
$J_{\rm MNS}^{(\mu,e)} = -A \sin \phi \equiv - J$.
Under the two-flavor framework,
this transition probability, $\nu_\mu^{}\to \nu_e^{}$ is written as
\begin{equation}
 P^{(2)}_{\nu_\mu \to \nu_e} =
\sin^2\theta_{\mu e} \sin^2\dfrac{\Delta_{\mu e}}{2}\,,
\eqlab{Pme20}
\end{equation}
where $\Delta_{\mu e}$
is the mass-squared difference
and
$\theta_{\mu e}$ is the unknown mixing angle.
We suppose that 
these two parameter, especially $\Delta_{\mu e}$,
are changed by the three-flavor effect.
As in the case of the survival probability,
we rewrite these parameters as
\begin{equation}
 \sin^2 \theta_{\mu e} = h\,,~~~~~
 \Delta_{\mu e} = \Delta_{13} + 2 \delta\,,
\eqlab{setH}
\end{equation}
where $\delta$ is smaller than $\Delta_{13}$
and $h$, in general, can take an arbitrary value.
Before estimating the three-flavor effect,
let us calculate the value of $h$ and $\delta$
for $\Delta_{12}=0$.
The transition probability becomes
\begin{equation}
P_{\nu_\mu^{} \to \nu_e^{}}=
4\left|\cU{\mu 3}{}\cU{e 3}{}\right|^2
\sin^2\dfrac{\Delta_{13}}{2}\,,
\eqlab{Pme00}
\end{equation}
with $\Delta_{12}=0$.
At the first peak of transition probability,
$\Delta_{\mu e}(E_{\rm peak})=\pi$,
\eqsref{Pme20} and\eqvref{Pme00} become
\begin{eqnarray}
P^{(2)}_{\nu_\mu \to \nu_e}
(\Delta_{\mu e} = \pi) &=&
h\,, \nn\\
P_{\nu_\mu^{} \to \nu_e^{}}
(\Delta_{\mu e} = \pi) &=&
4\left|\cU{\mu 3}{}\cU{e 3}{}\right|^2
\cos^2\delta\,,
\eqlab{Delta120}
\end{eqnarray}
respectively.
From these equations,
$h$ and $\delta$ are solved as
\begin{eqnarray}
h_0^{}
&\equiv&
h(\Delta_{12}=0,\Delta_{\mu e}=\pi)
=
4\left|\cU{\mu 3}{}\cU{e 3}{}\right|^2\,,\nn\\
\delta_0^{}
&=&\delta(\Delta_{12}=0,\Delta_{\mu e}=\pi)
=0\,,
\eqlab{h0}
\end{eqnarray}
when we suppose that $\Delta_{\mu e}$ is same as
$\Delta_{13}$ without three-flavor effect.

When we assume that $\Delta_{12}$ is nonvanishing, but small,
\eqref{Pme} becomes
\begin{eqnarray}
P_{\nu_\mu^{} \to \nu_e^{}}&=&
P_0(\mu,e)
+ P_1(\mu,e) \times \Delta_{12}
+ P_2(\mu,e) \times \Delta_{12}^{2}
+O(\Delta_{12}^3)\,,
\eqlab{Pme4}
\end{eqnarray}
where 
$P_0(\mu,e)$, $P_1(\mu,e)$ and $P_2(\mu,e)$ are 
\begin{subequations}
\begin{eqnarray}
P_0(\mu,e) &=&
4\left|\cU{\mu 3}{}\right|^2
 \left|\cU{e 3}{}\right|^2
\cos^2\delta^{} \nn\\
&=&
4\left|\cU{\mu 3}{}\right|^2
 \left|\cU{e 3}{}\right|^2 + O(\delta^2)\,,
\eqlab{Pme0d}\\
P_1(\mu,e) &=& 
2{\rm Re}\cA{23}{}
\sin 2\delta^{}
-
4J\cos^2\delta^{}\nn\\
&=&
4\delta{\rm Re}\cA{23}{}
-
4J+O(\delta^2)\,,
\eqlab{Pme1d}\\
P_2(\mu,e)&=&
-{\rm Re}\cA{12}{}
+{\rm Re}\cA{23}{}\cos{2\delta^{}}
+
J\sin2\delta^{}\nn\\
&=&
-{\rm Re}\cA{12}{}
+{\rm Re}\cA{23}{}
+2\delta J+O(\delta^2)\,,
\eqlab{Pme22}
\end{eqnarray}
\eqlab{Pmed_ex}
\end{subequations}
at the first peak $E_{\rm peak}$.
Here, the symbols $\cA{ij}{}$ are defined as
\begin{equation}
 \cA{ij}{} = \cU{\mu i}{} \cU{e i}{\ast} \cU{e j}{} \cU{\mu j}{\ast}\,.
\end{equation}
The value of $\cU{ij}{}$, $\cA{ij}{}$, and $\delta$
is fixed at $E_{\rm peak}$\footnote{{%
We drop here the label ``peak'' for
$\cU{}{\alpha j}$, $\cA{}{ij}$, and $\delta$,
for simplicity.
}}.
We assume that $h$ at $E_{\rm peak}$
is function of $\Delta_{12}$,
\begin{equation}
 h = h(\Delta_{12}) = h_0 + \sum_{k}a_k^{}\Delta_{12}^k\,,
\eqlab{defHF}
\end{equation}
where $h_0$ is equal to
$4\left|\cU{\mu 3}{}\cU{e 3}{}\right|^2$
and $a_k^{}$ are independent of the neutrino energy.
From \eqsref{Pmed_ex} and \eqvref{defHF},
$\delta$ can be solved as
\begin{equation}
\delta \simeq \dfrac
{
\left(a_1^{} + 4J\right)
 +
\left({\rm Re}\cA{12}{} - {\rm Re}\cA{23}{} + a_2^{}\right)\Delta_{12}^{}
+
a_3^{} \Delta_{12}^2
}
{
4{\rm Re}\cA{23}{} + 2 J \Delta_{12}^{}
}\,.
\eqlab{delta1}
\end{equation}
When we take the limit $\Delta_{12} \to 0$,
$\delta$ must vanish
because of \eqref{h0}.
Thus,
\begin{equation}
 a_1^{} = -4J\,,
\end{equation}
and 
\begin{equation}
\delta \simeq 
\dfrac{\left({\rm Re}\cA{12}{} - {\rm Re}\cA{23}{} + a_2^{}\right)
+a_3^{} \Delta_{12}}
{4{\rm Re}\cA{23}{} + 2 J \Delta_{12}^{}}
\Delta_{12}^{}
\,.
\eqlab{delta2}
\end{equation}
The denominator of \eqref{delta2} becomes 0
under some conditions that are related to the value
of the MNS matrix elements,
\begin{equation}
2{\rm Re}\hat\cA{23}{} +  \hat{J} \hat\Delta_{12}^{}
= 0\,,
\eqlab{deno1}
\end{equation}
where $\hat\cA{23}{}$, $\hat{J}$, and $\hat\Delta_{12}^{}$
denote some fixed value of them.
By using the mixing angles $\theta_{ij}$ and CP phase $\phi$,
$\hat\Delta_{12}^{}$ is written as
\begin{equation}
\hat\Delta_{12} = - \dfrac{2}{\sin \hat \phi}
\left(\cos \hat \phi - 
\tan \hat \theta_{12}
\tan \hat \theta_{23}
\sin \hat \theta_{13}
\right)\,.
\end{equation}
Since $\delta$ does not diverge at the first peak,
the numerator also has to be 0
\begin{equation}
\left({\rm Re}\hat\cA{12}{} - {\rm Re}\hat\cA{23}{} + a_2^{}\right)
+ a_3^{} \hat\Delta_{12} = 0 \,,
\eqlab{num1}
\end{equation}
under the same condition.
Using \eqref{deno1}, $a_2^{}$ and $a_3^{}$ become
\begin{equation}
 a_2^{} = -{\rm Re}\hat\cA{12}{}\,, 
~~~\mbox{ and }~~~
 a_3^{} = -\dfrac{1}{2} \hat{J}
\,. 
\eqlab{as} 
\end{equation}
Finally,
we obtain
\begin{equation}
 \delta(E_{\rm peak}) \simeq -
\dfrac{\Delta_{12}^{}}{4}
\dfrac{
2\left(
 {\rm Re}\hat\cA{12}{}
-{\rm Re}\cA{12}{}
+{\rm Re}\cA{23}{}
\right)
-\hat{J} \Delta_{12}}
{2{\rm Re}\cA{23}{} + J \Delta_{12}^{}}\,,
\eqlab{delta_a}
\end{equation}
and
\begin{equation}
 h(E_{\rm peak})
= 4\left|\cU{\mu 3}{}\right|^2 \left|\cU{e 3}{}\right|^2
 -4 A \sin\phi~ \Delta_{12}^{}
 -{\rm Re}\hat\cA{12}{}\Delta_{12}^2
  +O(\Delta_{12}^3)\,.
\eqlab{ha}
\end{equation}
Because the order of $\cA{ij}{}$ is less than 1,
the shift of the first peak for the transition probability
is not large.
\Eqref{ha} shows that 
the first-peak of the transition probability with
$\delta_{\rm MNS}=90^\circ$
is smaller than that with $\delta_{\rm MNS}=270^\circ$.
Since the value of Re$\hat \cA{12}{}$ is negative for 
$\rho \simeq 3.0 $ (g/cm$^3$),
the first-peak of transition probability for the CP-conserved case
is slightly larger than that with $\Delta_{12}=0$.
These features remain unchanged for the inverted hierarchy.
But the shifting direction of the first-peak is different
between the normal hierarchy and inverted one.

\section{summary}
In this article, we have estimated the three-flavor effect for 
the determination of the mass-squared difference and the mixing angle
from the survival and transition probability of $\nu_\mu^{}$.
From both probabilities,
the larger mass-squared difference is changed
by the three-flavor effect.
The order of magnitude of the difference between 
the larger mass-squared difference $\delta m^2_{13}$
and 
the mass-squared difference of the two-flavor analysis
$\delta_{\rm dip,peak}$
is not only proportional to $\delta m^2_{12}$ but also
to the MNS matrix elements. 
We also find the CP phase dependence for 
the transition probability.
If there is no three-flavor effect for $\nu_\mu$ survival and
transition probabilities, the first-dip energy of $\nu_\mu$ survival
probability and the first-peak energy of $\nu_\mu$ transition one are
as same as that from $\Delta_{13}$.
However,
each of them is different from the value of $\Delta_{13}$,
which are shown from \eqref{dip1} and \eqref{delta_a},
because of three-flavor effect.
This means that the value of $\delta m^2_{13}$ from $\nu_\mu$
survival probability is slightly different from that from
$\nu_\mu\to\nu_e$ transition one.
The order of them are not so smaller than that of the expected
error for the $\delta m^2_{13}$ in the future long base-line neutrino
oscillation experiment.
These results are useful to estimate the three-flavor 
effects for the value of the parameters
which are obtained from the numerical analysis.

{\it Acknowledgment}

I would like to thank 
F.~Borzumati,
K.~Hagiwara, 
K.~Nishikawa,
K.~Senda,
and 
T.~Takeuchi,
for useful discussions and comments.
I am also grateful to H.~Kai
for warmhearted supports.


\begin{thebibliography}{99}
\bibitem{atm}
IMB collaboration, (R.~Becker-Szendy \etal), \PR{D46}{3720}{1992}; SOUDAN2
  collaboration, (W.W.M.~Allison \etal), \PL{B391}{491}{1997}; hep-ex/9901024;
  The Super-Kamiokande collaboration (Y.~Ahie \etal) hep-ex/0404034.

\bibitem{K2K}
K2K collaboration (E.~Aliu, \etal), hep-ex/0411038.

\bibitem{solar}
Homestake collaboration, (B.T.~Cleveland \etal), Nucl.~Phys. (Proc.~Suppl.)
  {\bf B38}, 47 (1995); Ap. J. {\bf 496}, 505 (1998); Kamiokande collaboration,
  (Y.~Fukuda \etal), \PRL{77}{1683}{1996}; GALLEX collaboration, (W.~Hampel
  \etal), \PL{B477}{127}{1999}; SAGE collaboration, (J.N.~Abdurashitov \etal),
  \PR{C60}{055801}{1999}; Super-Kamiokande collaboration, (Y.~Fukuda \etal),
  \PRL{81}{1158}{1998}, \PRL{82}{1810}{1999}; SNO collaboration, (A.N.~Ahmed
  \etal), \PRL{92}{181301}{2004} and (A.W.P. Poon \etal),
  \EPJ{C33}{S823}{2004}.

\bibitem{KamLAND}
The KamLAND collaboration (T.~Araki, \etal), hep-ex/0406035.

\bibitem{T2K}
Y.Itow, \etal, hep-ex/0106019; The latest version is in {\sf
  http://neutrino.kek.jp/jhfnu/}.

\bibitem{nova}
I.~Ambats \etal, FERMILAB-PROPOSAL-0929.

\bibitem{BNL}
BNL Neutrino Working Group (M.~Diwan, \etal), hep-ex/0211001; M.~Diwan, \etal,
  \PR{D68}{012002}{2003}.

\bibitem{LSND}
LSND collaboration, \PRL{77}{3082}{1996} and \PRL{81}{1774}{1998}; KARMEN
  collaboration, \PRL{81}{520}{1998} and hep-ex/9809007.

\bibitem{miniboone}
The MiniBooNE collaboration (M.H.~Shaevitz), hep-ex/0407027.

\bibitem{nume}
\eg
G.L.~Fogli, E.~Lisi, D.~Montanino, A.~Palazzo,
\PR{D64}{093007}{2001};
J.N.~Bahcall, M.C.~Gonzalez-Garcia, C.~Pena-Garay
\JHEP{0108}{014}{2001};
I.~Mocioiu and R.~Shrock, \JHEP{0111}{050}{2001};
M.~Maltoni, T.~Schwetz, M.A.~Tortola, J.W.F.~Valle, \PR{D68}{113010}{2003}
G.L.~Fogli, E.~Lisi, A.~Marrone, D.~Montanino, A.~Palazzo,
A.M.~Rotunno, hep-ph/0310012;
M.~Maltoni, T.~Schwetz, M.A.~Tortola, J.W.F.~Valle,
New J.Phys. {\bf 6} 122 (2004);
M.C.~Gonzalez-Garcia, M.~Maltoni, hep-ph/0406056;
J.N.~Bahcall, M.C.~Gonzalez-Garcia, C.~Pena-Garay, \JHEP{0408}{016}{2004};
S.~Goswami, A.~Bandyopadhyay, S.~Choubey, hep-ph/0409224;
M.C.~Gonzalez-Garcia, hep-ph/0410030;
W.~Winter, hep-ph/0410354,
and references therein.


\bibitem{MNS}
M.~Nakagawa Z.~Maki and S.~Sakata, \PTP{28}{870}{1962}.

\bibitem{PDB}
B.~Kayser, in Review of Particle Physics, \PL{B592}{145}{2004}, and references
  therein.

\bibitem{CKM}
N.~Cabibbo, \PRL{10}{531}{1964}; M.~Kobayashi and T.~Maskawa
  \PTP{49}{652}{1973}.

\bibitem{HO1}
K.~Hagiwara and N.~Okamura, \NP{B548}{60}{1999}.

\bibitem{H2B}
M.~Aoki, K.~Hagiwara, Y.~Hayato, T.~Kobayashi, T.~Nakaya,
K.~Nishikawa, and N.~Okamura, \PR{D67}{093004}{2003}.

\bibitem{CHOOZ}
The CHOOZ collaboration (M.~Apollonio, \etal),
\EPJ{C27}{331}{2003}, hep-ex/0301017.


\bibitem{JarP}
C.~Jarlskog.
\newblock \PRL{55}{1039}{1985} and Z. Phys. {\bf C29}, 491 (1985).

\bibitem{MSW}
L.~Wolfenstein, \PR{D17}{2369}{1978}; S.P.~Mikheyev and A.Yu.~Smirnov, Yad.
  Fiz. {\bf 42}, 1441 (1985) [Sov.J.Nucl.Phys.{\bf 42}, 913 (1986)] and Nuovo
  Cimento {\bf C9}, 17 (1986).
\end{thebibliography}
\end{document}